# Cavity-Enhanced Linear Dichroism in a van der Waals Antiferromagnet

Huiqin Zhang,[1,∥] Zhuoliang Ni,[2,∥] Aofeng Bai,[3] Frank Peiris,[3] Liang Wu,[2] Deep Jariwala[1]*

[1] *Department of Electrical and Systems Engineering, University of Pennsylvania, Philadelphia, PA 19104, USA*

[2] *Department of Physics and Astronomy, University of Pennsylvania, Philadelphia, PA 19104, USA*

[3] *Department of Physics, Kenyon College, Gambier, OH 43022, USA*

[∥]*These authors contributed equally: Huiqin Zhang, Zhuoliang Ni*

*Corresponding author: dmj@seas.upenn.edu

**Abstract:**
**Optical birefringence is a fundamental optical property of crystals widely used for filtering and beam splitting of photons. Birefringent crystals concurrently possess the property of linear dichroism (LD) that allows asymmetric propagation or attenuation of light with two different polarizations. This property of LD has been widely studied from small molecules to polymers and crystals but has rarely been engineered per will. Here, we use the newly discovered spin-charge coupling in van der Waals antiferromagnetic (AFM) insulator $FePS_3$ to induce large in-plane optical anisotropy and consequently LD. We report that the LD in this AFM insulator is tunable both spectrally and magnitude wise as a function of cavity coupling. We demonstrate near-unity LD in the visible-near infrared range in cavity-coupled $FePS_3$ crystals and derive its dispersion as a function of cavity length and $FePS_3$ thickness. Our results hold wide implications for use of cavity tuned LD as a diagnostic probe for strongly correlated quantum materials as well as opens new opportunities for miniaturized, on-chip beam-splitters and tunable filters.**

**Introduction:**

Optical anisotropy plays a crucial role in light manipulation for photonic and optoelectronic applications. Although giant out-of-plane optical anisotropy has been recently observed in the visible to mid-infrared wavelength range in perovskite chalcogenides, h-BN crystals and Mo, W chalcogenides[1], the in-plane optical anisotropy has been scarcely explored in low-dimensional van der Waals crystals. This has inspired an ongoing search for giant in-plane optical anisotropy among natural and artificial materials. To achieve in-plane anisotropy a material must possess some asymmetry in structural or electronic order along one direction. Often, this is achieved via anisotropic, one-dimensional (1D) structure of the material itself such as in liquid crystals or polymers[2] or carbon nanotubes. An effective way to quantify this anisotropy is via means of linearly polarized light spectroscopy often used to obtain linear dichroism (LD, the difference in reflected light intensities along two different polarizations)[1] or birefringence (the difference in refractive indices (△n) along two different polarizations), which represent a measure of anisotropic optical response. The property of LD has been very useful in producing versatile optical components[3, 4], including polarizers, wave plates, mirrors, and phase-matching elements. Despite its great fundamental importance and technological significance, LD and in-plane LD in novel materials is rarely studied systematically, and further, even little has been known about tuning or engineering LD in low-dimensional optical materials. Among known in-plane anisotropic materials such as Black Phosphorous (BP) and ReS$_2$ LD has been recently measured with magnitudes of ~ 20% to 40%[5, 6] which primarily emanates from the in-plane asymmetry of atomic arrangements or crystal structure. However, these are static values and no investigations on engineering/tuning them have been reported thus far. A notable approach to tune/engineer optical anisotropy involves the use of artificially designed metamaterials and metasurfaces which comprise of patterned nanostructures and have been shown to exhibit large birefringence values[7]. However, the widespread usage of such structures is impeded by optical losses and micro-nano fabrication challenges particularly in the visible (VIS) to near-infrared (NIR) domain. Therefore, an in-depth study of LD and birefringence, particularly the ability to tune and enhance LD, is notably lacking. This ability to tune the LD would be particularly vital for wavelength tunable beam-splitters, waveguides and detectors.

The recent emergence of van der Waals chalcogenides with long-range magnetic order offer a unique opportunity in tuning light-matter interactions as a function of magnetic phase transitions or ordering. Magnetic phase transitions can induce spin-charge couplings that can result in breaking the symmetry of optical response in-plane leading to observation of LD. Magneto-optical effects such as magneto-optical Kerr effect (MOKE)[8, 9] are commonly used to detect such magnetic phase transitions and also spin configurations or ordering in ferromagnets. As compared to ferromagnets, however, the detection of antiferromagnetism (AFM) is much more difficult due to the lack of a net magnetic moment. Prior works have used neutron scattering[10, 11], Raman spectroscopy[12, 13, 14, 15] and second harmonic generation[16, 17, 18] to detect the symmetry of spin ordering in antiferromagnets(AFM), based on the expansion of the unit cell, magnetoelastic coupling, and inversion symmetry breaking, respectively. Here we show that in van der Waals AFM insulator FePS$_3$ a strong spin-charge coupling can induce large in-plane anisotropy leading to extraordinarily high LD values. These LD values can be further tuned by coupling them with a simple optical cavity medium which results in enhancement of the LD to near-unity values. In addition, the tuning of the cavity further allows spectral tuning of this large LD response. Our results suggest that van der Waals AFM semiconductors and insulators are outstanding candidates for the tuning and amplification of in plane LD in the visible to NIR range and open the door to novel, multi-spectral, miniature nanophotonic components by virtue of spin-charge coupling phenomena.

**Results:**

We report the observation of near-unity, tunable and cavity-enhanced linear dichroism in a broad-band visible-NIR region of the spectrum in thin layers of vdW AFM material FePS$_3$. FePS$_3$ belongs to a class of transition metal phosphorous trichalcogenides (MPX$_3$, M: Fe, Mn, Ni and X: S, Se), which are van der Waals AFM insulator materials with bandgap ranging from 1.3eV to 3.5eV[12, 13, 19, 20, 21]. Within individual FePS$_3$ layers, Fe atoms are arranged in a honeycomb lattice structure. The spins pointing out of plane along a chosen row of Fe atoms aligned in the zigzag direction (shown along the x-axis) are opposite to the spins in the adjacent spin chain of Fe atoms (Fig. 1a, the in-plane structure), forming a zigzag AFM order[22]. The blue and red Fe atoms are aligned at +z and -z direction, respectively. The interlayer coupling between adjacent layers is AFM, as shown in figure 1.a (out-of-plane structure). Raman spectroscopy[12] and magnetic susceptibility[23] measurements from prior literature have shown that the Néel temperature is ~ 118 K in bulk FePS$_3$. Below Néel temperature, the strong spin-charge correlation leads high optical anisotropy which is

locked into the zigzag direction of FePS$_3$ crystal.[24]

To demonstrate this, linearly polarized optical reflectance measurements were performed on multilayer FePS$_3$, directly exfoliated on a Si/SiO$_2$ substrate (figure 1.b, SiO$_2$: 90nm thickness). The samples show varying colors with varying thicknesses suggesting strong optical interference effects. We use a polarizer placed in the reflected light path to the detector to detect the polarized reflectance and the linear dichroism (LD) from the sample (figure 1.c). The axes along which in-plane anisotropy is induced are also shown in figure 1.b. Within this multilayer FePS$_3$ structure, the LD can be enhanced from the cavities formed vertically inside the material which is indicated by the orange oval marked in figure 1.c.

We define LD as $\frac{R_\perp - R_\parallel}{R_\perp + R_\parallel}$, where $R_\parallel$ ($R_\perp$) is the peak intensity of horizontally (vertically) polarized optical reflection (see methods)[25]. The above expression can also be understood as difference in reflection between the two axes divided by total reflect unpolarized reflection. Figure 1.d shows the high degree of optical anisotropy with LD up to 98.6% for measurements performed at 30 K at 751 nm wavelength. The polar plot indicates the spin chain direction of the measured sample (aligned as Rx and Ry in figure 1.b). The LD is observed to be a strong function of the temperature as shown in Figure 1.e. The transition of LD from ~10% to ~100 % begins at 130 K and is complete by 110 K coinciding with a precipitous drop at the Néel temperature (T$_C$ ~118 K). The LD magnitude approaches the highest level of 98.6% below Néel temperature, following the antiferromagnetic phase transition. This temperature dependence is similar to prior reports on temperature dependent Raman spectroscopy[12, 13] and magnetic susceptibility[23] which indicates its spin-charge coupling due to the AFM ordering. This is clear evidence of LD induced by the AFM transition and spin ordering along the zigzag crystallographic direction.

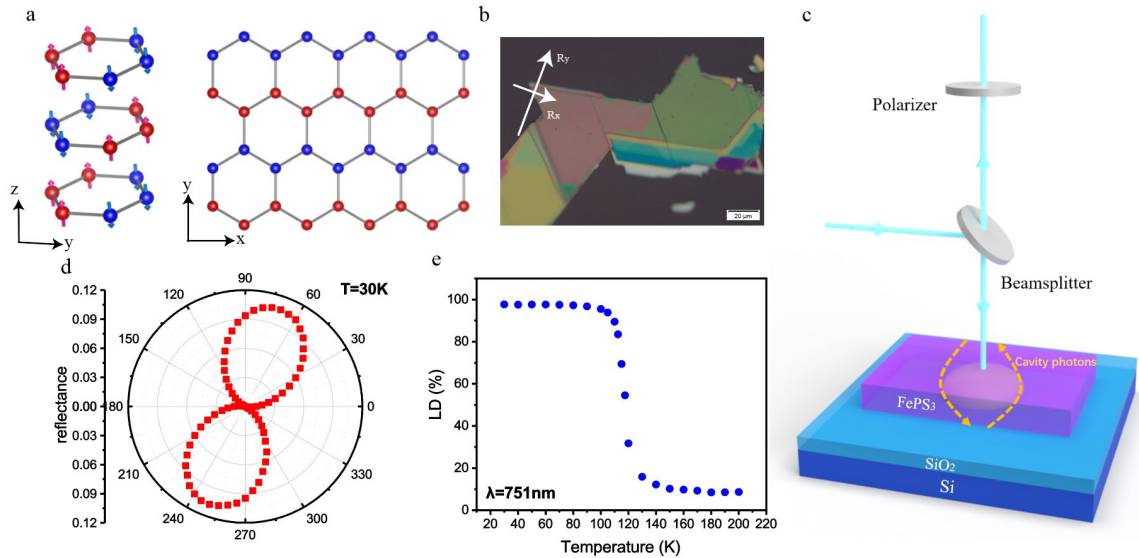

**Figure 1 Optical in-plane anisotropic behavior in AFM van der Waals FePS$_3$.** a. Schematic of zigzag AFM order in a single layer FePS$_3$ (right) and in multilayer stacks (left). The blue and red atoms represent the Fe atoms with spins in opposite directions. b. Optical image of exfoliated FePS$_3$ on Si/SiO$_2$ substrate indicating the crystallographic axes (x and y corresponding to the ZZ and AC direction respectively). c. Schematic of the experimental setup of the linearly polarized optical

reflectance microscope used to measure the multilayer FePS$_3$. The cavity photons (indicated by an orange oval) form inside the layered material. d. Polar plot of FePS$_3$ reflectance intensity as a function of linear polarization angle. Data is taken at 30 K with an incident light wavelength at 751nm. e. The linear dichroism at 751 nm of FePS$_3$ multilayer sample as a function of temperature showing T$_C$=118 K.

To further understand the nature of this optical response, we perform polarized spectral reflectance in the VIS-NIR range and estimate the LD spectrum. Figure 2.a shows an optical micrograph and the corresponding AFM topography of the FePS$_3$ crystal sample. The AFM line cut shows the step edge of the sample corresponding to a thickness of 268nm. As shown in the polarization angle-dependent reflectance spectra (figure 2.b), the extinction (reflectance) is highly dependent on the incident light polarization and consequently also the LD (Figure 2c). Several resonant absorption (reflection) peaks are visible with values corresponding to near-unity absorption, particularly at ~750 nm. These resonant reflectance peaks suggest that the reflected light is strongly attenuated in the layered material due to the thin film interference effect[26]. Briefly, a film with a thickness of $k\frac{\lambda}{2n_{eff}}$ (k is an integer as an order number, λ is the incident light wavelength, n$_{eff}$ is the effective refractive index of the film in the specific structure) would form an optical cavity where light (photons) of specific wavelengths will be trapped inside the film and thereby result in enhancement of the absorption[27, 28]. Hence, the resonant absorption (reflectance) peaks in figure 2.b correspond to different orders of Fabry-Perot cavity mode resonances. The high anisotropy is also observed in the polarized reflectance spectra. The absorption resonances significantly vary at different polarized angles illustrating the in-plane birefringence (Δn) of FePS$_3$ since the cavity resonance mode is highly dependent on n$_{eff}$. The lattice anisotropy axes can be determined by plotting the polar plot of the reflectance intensities which also depends on the wavelength, as is shown in figure 2.c. The polarization angle of 340° (70°) corresponds to the long (short) axis of lattice anisotropy. We also perform temperature dependence reflectance measurements along the long (340°, figure 2.d) and short axes (70°, figure 2.e). It is evident from these reflectance measurements that the cavity resonance at 765 nm blue-shifts (red-shifts) while the cavity resonance at 523 nm red-shifts (blue-shifts) at 340° (70°) polarized angle as FePS$_3$ goes through AFM ordering below its transition temperature. This indicates that the birefringence originates from the AFM order, and the birefringence is also spectrally dispersive. The reflectance of at 751 nm (862 nm) is the highest (lowest) at 340° polarization angle while it is lowest (highest) at 70° polarization. Figure 2.f shows the temperature dependence of LD spectrum as derived from Figure 2.b. It is evident from the plot that the LD intensity variations extend broadly in the visible and NIR spectrum. Multiple peaks are observed in the LD spectrum corresponding to multiple cavity modes formed by the crystal in the reflectance spectrum in figure 2.b. The largest LD peak reaches up to 98.6% corresponding to the near-unity cavity-resonant absorption peak at 751 nm and the second-largest LD reaches to 80.0% corresponding to absorption peak at 862 nm in the reflectance spectrum. More LD peaks at shorter wavelengths show the lower intensity and correspond to the higher-order resonant-cavity modes. The absorption losses at these cavity-resonances correspond to absorption in both the FePS$_3$ as well as the underlying Si/SiO$_2$. The temperature dependence of LD spectra shows a drastic reduction in the magnitude of the LD above 118 K, matching well with the Néel temperature of FePS$_3$. More reflectance and LD measurements have been performed on multiple samples with varying thicknesses ranging from 80 nm to 270 nm shown in Supplementary Information. The cavity-enhanced LD are observed similarly in various flakes, and multiple LD peaks are also observed at different wavelength ranges due to the various thicknesses of the samples (See supporting information S1 for more details).

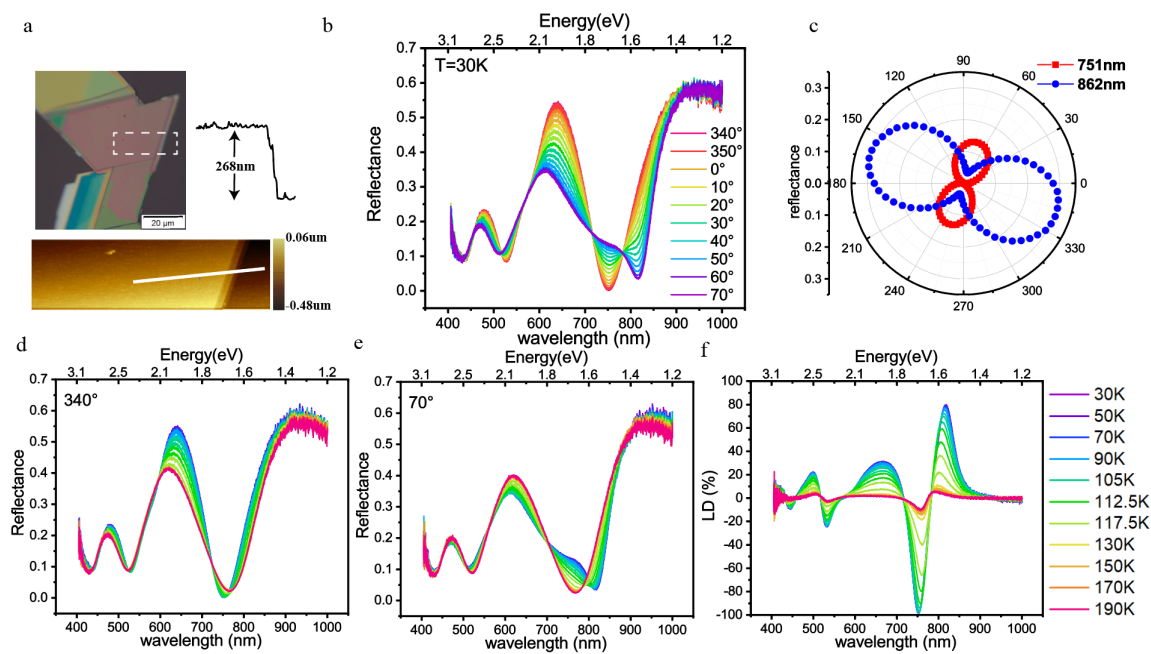

**Figure 2 Linear dichroism spectrum of FePS$_3$.** a. Optical image of the exfoliated FePS$_3$ crystal sample. Inset shows an atomic force topography image of the area enclosed by the dashed rectangle. Line cut (white line) and the corresponding height plot indicates the sample thickness as ~268 nm. b. Polarization angle dependence of reflectance intensity spectra measured at 30 K with a step of 5 degrees polarization rotation from 340-70 degrees. c. Polar plot of FePS$_3$ reflectance intensity as a function of linear polarization angle (steps of 5 degrees) at a specific wavelength of 751 nm (red) and 862 nm (blue) respectively. The large anisotropy of in-plane crystallographic axes is evident. d,e Temperature dependence of the reflectance intensity spectra at polarization angles of 340° (d) and 70°(e). f. Temperature dependence (30 K to 200 K) of the linear dichroism (LD) spectrum.

To better understand the cavity-enhanced LD of this structure and its thickness dependence, we perform electromagnetic wave calculations using the transfer matrix method (TMM) to estimate the thickness-dependent reflectance and the LD intensity (See supplementary Information S2 for calculation methods and related details). To make a correct estimate from TMM calculations we measure the dielectric function of FePS$_3$ by spectroscopic ellipsometry (See Supplementary Information 3). In addition, an estimate of the birefringence at 30K is also needed for accurately calculating the LD. Polarization-dependent reflectance measurements on bulk samples were performed for estimating the bulk crystal LD, (see Supplementary Figure 4.a). As seen in Figure 3.a the bulk LD spectrum of FePS$_3$ measured at 30 K is much less dispersive with the largest magnitude of LD at ~ 8%. To estimate the extent of birefringence (Δn) in the bulk samples, the following relation $n_{1,2} = \frac{1+\sqrt{R_{1,2}}}{1-\sqrt{R_{1,2}}}$ [29] is applied to the reflectance spectrum and quantify the refractive indices along the two axes (n1, n2) which represent the highest degree of anisotropy. This quantification at every wavelength result in the estimation of the birefringence (Δn) spectrum (figure 3.b). The birefringence ranges from -0.2 to 0.075 depending on the wavelength. The birefringence spectrum shows an interesting feature of crossing over from positive values below 650 nm to negative values at above 650 nm, indicating the presence of an electronic band resonance in FePS$_3$ [30]. This behavior also explains the opposite directions of shifts in cavity-resonance peaks at 523 nm and 765 nm position respectively in the reflectance spectra shown in Figure 2.d & e. Additional details on the estimation of the birefringence spectra are provided in Supplementary Information 4. The real part of the refractive index (n) in the visible range as estimated from ellipsometry is also shown in the inset of Figure 3.b. Knowing the birefringence values as a function of wavelength, the dielectric function from ellipsometry and further assuming that the refractive indices along the two axes of maximum anisotropy are $n1 = n - \frac{\Delta n}{2}$, $n2 = n + \frac{\Delta n}{2}$ respectively, one can perform TMM calculations and plot the layer-dependent reflectance as well as LD (See supporting information 4). Figure 3.c shows the experimental LD spectra of FePS$_3$ for various thicknesses. The corresponding simulated LD spectra (Figure 3.d) show a close qualitative match in terms of resonances and features observed in the experimental spectra as indicated by the corresponding labels. There is, however, some quantitative mismatch between the simulation and the experimental result due to two main reasons: First, the estimation of the birefringence is not perfect in this case since we neglect the dielectric loss at low temperature. Second, the AFM domains of FePS$_3$ may not be the same size and ideally matched in each layer, which means the sampled thickness in the experiment may not be the ideal thickness for the calculations. To further understand the nature of these resonances, we also check the electric field distributions along the cross-section of FePS$_3$ crystal, i.e. in the plane normal to the basal plane of the crystal, as shown in figure 3.e. For thin-film interference, the cavity resonance occurs at $d = k\frac{\lambda}{2n_{eff}}$ (where k is the order of the mode) which is the approximate value of the resonant wavelength at the specific thickness (Figure 3.c &d). The peaks in LD are labeled with an upright triangle, a square and a star symbols (Figures 3.c-e) and correspond to first, second and third-order cavity modes respectively formed inside FePS$_3$ crystal (in 3.e). The field profiles again verify that the LD is enhanced due to the cavity modes.

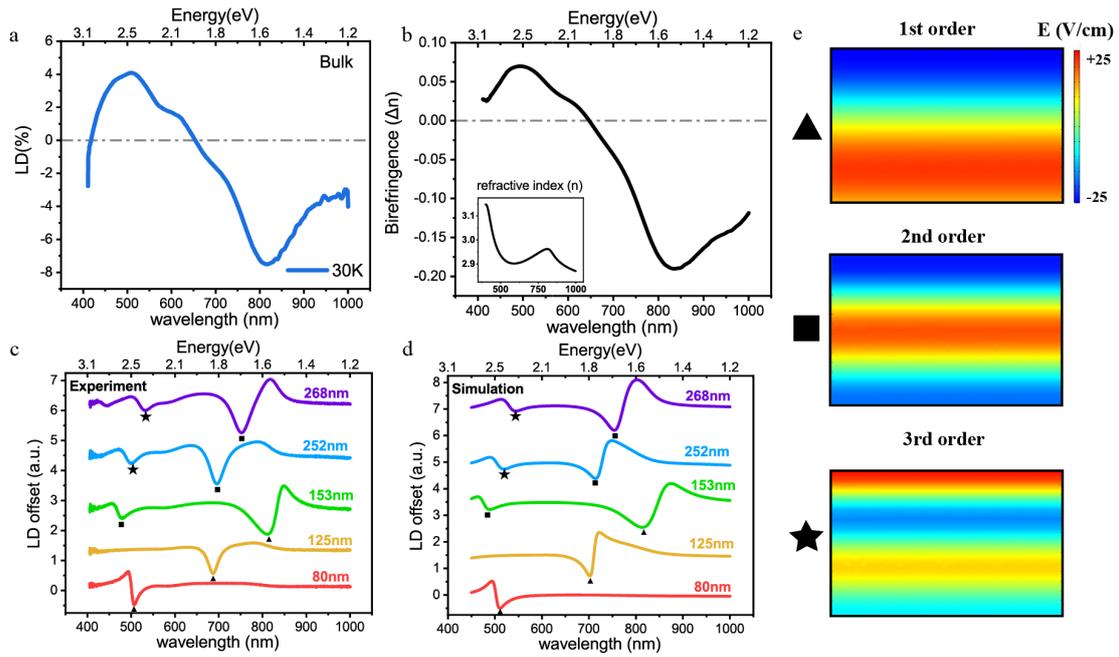

**Figure 3 Simulation model of cavity enhanced multilayer FePS$_3$.** a. LD spectrum of bulk (> 100 microns thickness) FePS$_3$ crystal measured at 30 K. b. The birefringence spectrum at 30 K calculated from (a). The spectrum shows a wavelength dependent birefringence of FePS$_3$, indicating a positive Δn below 650 nm which transitions to a negative Δn above 650 nm. The inset is refractive index (real part) of FePS$_3$ measured via spectroscopic ellipsometry method (298 K). c. Experimental LD spectra for specific FePS$_3$ thicknesses of 80 nm, 125 nm, 153 nm, 252 nm, 268 nm. All measurements were performed at 30 K. d. Transfer matrix calculated LD spectra for the same thicknesses as in (c). The LD peaks match well in terms of qualitative shape and spectral positions. e. Electric-Field profiles of the Fabry-Perot cavity-mode resonance peaks labelled in (c) and (d). The triangles indicate the first-order cavity mode, while the squares and the stars indicate the second-order and third-order cavity modes respectively.

Given that the resonance magnitude and spectral position in LD is tunable as a function of FePS$_3$ thickness, this system provides a unique opportunity in exploring tunability of the LD spectrum. This can be done in two ways. Figure 4.a shows the approaches adopted here to tune the optical cavity sizes using both internal and external cavity media. For a fixed thickness of the substrate layer (i.e. bulk Si with some fixed SiO$_2$ thickness on top), if the thickness of FePS$_3$ is much smaller (4.a (i)) than the incident wavelength, then the light will pass through FePS$_3$, leaving the material with minimal loss. Likewise, if the thickness is much larger (4.a (ii)) than the light will be fully absorbed or reflected before reaching the bottom-most layers of FePS$_3$. In neither of the two cases will the FePS$_3$ layers form and support cavity modes. Only when the thickness of FePS$_3$ is at a comparable level to that of the incident light wavelength (4.a (iii)), the material will be capable of supporting a stable cavity for the photons, forming a standing wave inside it. In this case, the photons are trapped in the FePS$_3$, which enhances the light-matter interaction and thus the LD. Similarly, one can achieve the spectral tuning of the LD peaks by changing the size of external cavities (SiO$_2$, 4.a (iv)). The detailed electric field profiles of the cavity resonances formed by the internal cavity (FePS$_3$) and external cavity (SiO$_2$) are shown in comparison in Supplementary Information 5. First, by sweeping the thickness of FePS$_3$, one can simply achieve the spectral tuning of the LD peaks by tuning the cavity size itself. Figure 4.b shows a color map of LD spectra as a function of FePS$_3$ thickness with the LD magnitude and sign (red positive, blue negative) shown on the color scale. The stars, squares and triangular symbols correspond to the experimentally observed LD peaks in figure 3(c). In this map, the linearly dispersive branches of the LD resonances correspond to the first, second and third-order Fabry-Perot cavity modes respectively. The lowest order modes show the strongest LD contrast while the higher order modes show weaker contrast. The color plot further shows that the desired thickness range for FePS$_3$ crystals on 90 nm SiO$_2$/Si is ~270 nm for observation of maximum magnitude of LD under reflectance in the visible wavelengths range. The second way of tuning the LD is to tune the substrate SiO$_2$ thickness. For a fixed layer thickness of FePS$_3$ below 50 nm where the material is too thin to form any internal cavity to support a mode, we can use the external cavity (SiO$_2$) to enhance the absorption and LD of FePS$_3$ as seen in Figure 4.c. Here, the thickness of FePS$_3$ layer is fixed at ~15 nm. Multiple orders of LD peak branches are observed and the symbols in the spectral map once again denote the experimental LD peak positions of 1$^{st}$ order (triangles), 2$^{nd}$ order (square) and 3$^{rd}$ order (star) resonances. The corresponding full experimental spectra (line-plots) are shown in Supplementary Information 6. Several FePS$_3$ crystal samples with roughly the same thickness (~15 nm) when measured on Si substrates with different SiO$_2$ thicknesses (50 nm, 90 nm, 280 nm) show shifting of the LD resonance peak (Figure 4.c). This observation further verifies that the LD response is tunable by changing the optical cavity sizes.

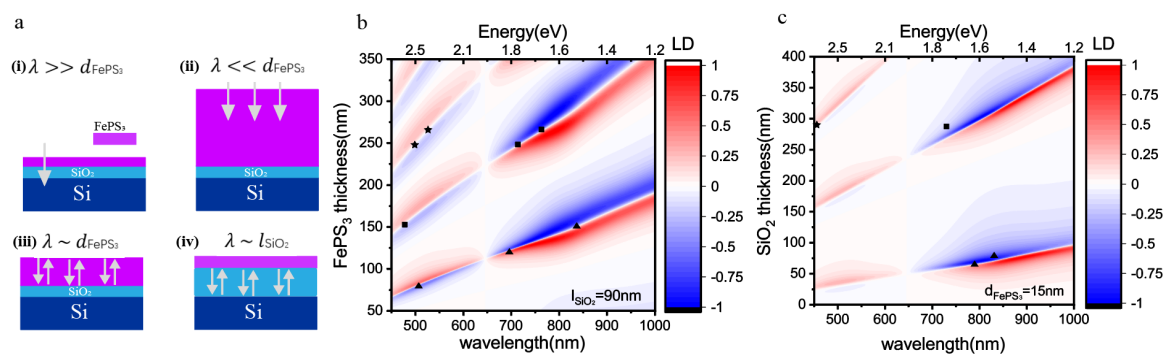

**Figure 4 Spectral tuning of LD enhancement by tuning cavity sizes.** a. Schematic illustration of internal and external Fabry-Perot cavity mode formation in Si/SiO$_2$/FePS$_3$ stacks with varying thicknesses of FePS$_3$ (internal cavity) and SiO$_2$ (external cavity). Resonant enhancement of the LD signal is observed at the cavity modes. b. Calculated LD spectral color plot as a function of FePS$_3$ thicknesses. The color scale on the map indicates the magnitude of LD resonance (red positive, blue negative) induced by the internal cavity formed in FePS$_3$ which is tuned by changing the FePS$_3$ thickness. The black symbols correspond to the experimentally measured LD peaks from figure 3(c). c. Calculated LD spectral color plot as a function of varying dielectric layer SiO$_2$ thicknesses. The map indicates that the LD resonances in thin FePS$_3$ can be spectrally tuned by the external cavity formed by varying index between Si/SiO$_2$/FePS3 and can be tuned simply by changing the SiO$_2$ thickness upon keeping FePS3 thickness constant at a thin value. The black symbols correspond to the experimentally measured 1$^{st}$ order (triangle), 2$^{nd}$ order (square) and 3$^{rd}$ order (star) LD resonances, respectively.

**Conclusion:**

The experiments, calculations and results above have several important implications both at a fundamental level and in terms of potential for future applications. First of all, LD spectroscopy is a valuable technique in probing anisotropic spin structures in van der Waals AFM materials. Second, since the optical anisotropy axis is locked to the zigzag direction in real space along which the spin chains are aligned, polarized LD can also serve as a measure or a probe in detecting or imaging AFM domains. Finally, the implications on the applied side directly emanate from the observation of spectrally tunable LD with very high magnitudes. Under reflection mode measurement, the LD observed is near-unity in magnitude and is tunable across the entire visible spectrum either by changing the $FePS_3$ or the $SiO_2$ thickness. This magnitude and tunability of the LD is remarkable and higher than the observed LD in all other known anisotropic 2D materials, ($GeSe$[31], b-AS[32], $ReS_2$[6], BP[5], α-$MoO_3$[33], GaTe[34], TiSe[35], $BaTiS_3$[36].) as seen in figure 5. Upon performing the same LD analysis as the one done in this work one can see that the magnitude of LD just ranges from 5% to 60% in these other in-plane anisotropic crystals. Here by virtue of cavity coupling and engineering the stack structure, we have been able to show enhanced and spectrally tunable LD that coincides with the cavity resonances.

The ability to tune LD both in magnitude and spectrally via artificial design of meta-structures and materials stack opens new opportunities in nanophotonic component design particularly the use of ultrathin and highly birefringent materials for on-chip filtering, beam splitting as well as non-linear optical components.

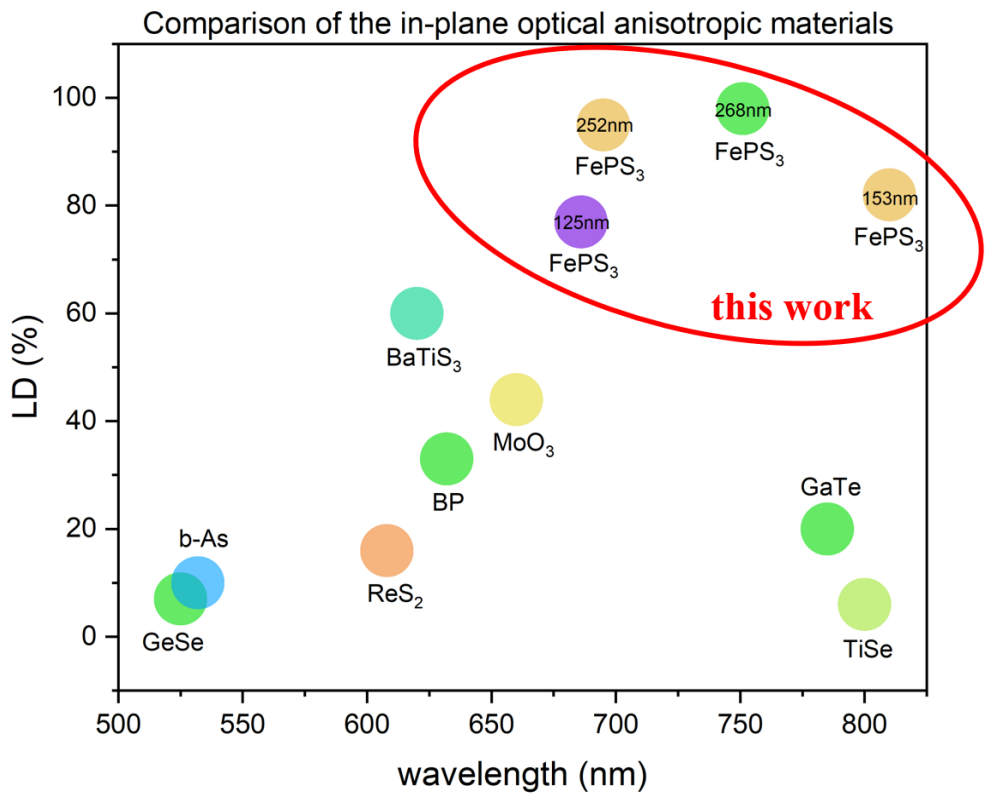

**Figure 5 Comparison of linear dichroism in in-plane optically anisotropic materials with observations in $FePS_3$ in this work.** The low dimensional materials with known in-plane optical anisotropy in the visible light regime ( $GeSe$[31], b-AS[32], $ReS_2$[6], BP[5], α-$MoO_3$[33], GaTe[34], TiSe[35], $BaTiS_3$[36])

are shown with colored circles and respective wavelength of the measurements. The LD ranges from 5% to 60% for most cases, far lower than multilayer $FePS_3$ as reported in this work. The numbers in the $FePS_3$ labelled circles denote the crystal thicknesses which illustrates the spectral and magnitude tunability of LD for different cavity sizes.

## Methods

### Sample preparation

$FePS_3$ layers were mechanically exfoliated from bulk crystal (HQ-graphene) using Scotch Tape and transferred onto the $Si/SiO_2$ substrate with thermally grown oxide in a dry process. The $Si/SiO_2$ wafers with $SiO_2$ thicknesses of 50 nm, 90 nm and 280 nm, were commercially purchased. For sample thickness-dependent measurements, samples are transferred on the same substrate with $SiO_2$ thickness of 90 nm. For dielectric layer-dependent measurements, different $FePS_3$ crystal samples with the same thickness of ~15 nm were transferred onto the substrate with 50 nm, 90 nm, 280 nm $SiO_2$ thickness.

### Reflectance measurement

The normalized reflectance intensity from 0 to 1 are all achieved by subtracting the background and normalizing with a reflectance intensity from a silver mirror.

$$R_{normalized} = \frac{R_{sample} - R_{background}}{R_{silver}}$$

### Linear dichroism measurement

An unpolarized halogen light source (AvaLight-HAL) is focused on the sample by a 50X objective (Olympus SLMPLN 50X N.A.=0.35). The reflected light is analyzed by a linear polarizer and collected by a multi-mode optical fiber into a spectrometer with a charge-coupled device camera. The polarization-dependent LD spectrum is calculated by $\frac{R_\perp - R_\parallel}{R_\perp + R_\parallel}$, where $R_\perp$, $R_\parallel$ are the reflectance perpendicular and parallel to the polarization[25]. The sample was loaded in a helium-free cryostat for these measurements.

### Ellipsometry

Ellipsometry spectra were obtained at three angles of incidence (65°, 70° and 75°) using a VASE ellipsometer (J.A. Woollam). Since the sample size was small, a focusing-optics module with a spot size of ~100 μm was attached to the instrument. Additionally, the instrument was equipped with an auto-retarder to monitor the depolarization caused mainly by non-specular reflections[37]. Scans with high depolarization values (~10% or higher) were discarded as they were harder to model, and scans with only low depolarization values were used obtain the dielectric function of the sample. More details of the dielectric function results are in Supplementary Information Part 2.

### Calculations

All the calculations and simulations are done by the transfer matrix method. Details are discussed in Supplementary information Part 4.

**Contributions**
D.J., H.Z. and Z.N. conceived the project. H.Z. and Z.N. made the samples, performed linear polarized reflectance measurements, and atomic force microscopy characterization. H.Z. and Z.N. performed the calculation work. F.P. and A.B performed the ellipsometry measurements. H.Z. analyzed and interpreted the optical spectroscopy and simulation data with help from Z.N. and D.J. H.Z. and D.J. wrote the paper with input from all co-authors. D.J. supervised the study.

**Acknowledgements**
D.J. acknowledges primary support for this work by the U.S. Army Research Office under contract number W911NF-19-1-0109. D.J. also acknowledges support from National Science Foundation (NSF) supported University of Pennsylvania Materials Research Science and Engineering Center (MRSEC) (DMR-1720530). H.Z. was supported by Vagelos Institute of Energy Science and Technology graduate fellowship. L.W. acknowledges support from the ARO under the Grants W911NF1910342 and W911NF2020166 and a seed grant from NSF supported University of Pennsylvania Materials Research Science and Engineering Center (MRSEC) (DMR-1720530). F.P. acknowledges support from Kenyon College and NSF grant DMR-2004812. The authors acknowledge assistance from Jason Lynch for spectroscopic ellipsometry measurements.

**Conflicts of interests:** The authors declare no competing or conflicting interests.

**Data availability:** The data that support the conclusions of this study are available from the corresponding author upon request.

**Code availability:** The codes used in this study for plotting and modelling are available from the corresponding author upon request.